\documentclass[sts]{imsart}

\RequirePackage[OT1]{fontenc}
\usepackage{amsthm,amsmath,amsfonts,comment}
\usepackage[numbers]{natbib}
\usepackage{bm}
\RequirePackage{graphicx,xcolor}

\startlocaldefs
\numberwithin{equation}{section}
\endlocaldefs

\newcommand{\var}[1]{\textrm{var}\left({#1}\right)}
\renewcommand{\vec}[1]{\bm{#1}}
\newcommand{\vecX}{\vec{X}}
\newcommand{\vecY}{\vec{Y}}
\newcommand{\veps}{\vec{\epsilon}}
\newcommand{\veta}{\vec{\eta}}
\newcommand{\vtheta}{\vec{\theta}}
\newcommand{\vphi}{\vec{\phi}}

\newcommand{\flu}{A/H1N1pdm influenza}

\begin{document}

\begin{frontmatter}
\title{Evidence synthesis for stochastic epidemic models}
\runtitle{Evidence Synthesis for Stochastic Epidemic Models}

\begin{aug}
\author{\fnms{Paul J} \snm{Birrell}%
\ead[label=e1]{paul.birrell@mrc-bsu.cam.ac.uk}},
\author{\fnms{Daniela} \snm{De Angelis}\corref{}\ead[label=e2]{daniela.deangelis@mrc-bsu.cam.ac.uk}}
\and
\author{\fnms{Anne M} \snm{Presanis}
\ead[label=e3]{anne.presanis@mrc-bsu.cam.ac.uk}
\ead[label=u1,url]{www.foo.com}}

\runauthor{Birrell, De Angelis \& Presanis}

\affiliation{MRC Biostatistics Unit, Cambridge}

\address{Paul Birrell is a Senior Investigator Statistician, Daniela De Angelis \printead{e2} is a Programme Leader and Anne Presanis is a Senior Investigator Statistician all at the MRC Biostatistics Unit, Cambridge Biomedical Campus, Cambridge Institute of Public Health, Forvie Site, Robinson Way, Cambridge CB2 0SR, UK.}

\end{aug}

\begin{abstract}
In recent years the role of epidemic models in informing public health policies has progressively grown. Models have become increasingly realistic and more complex, requiring the use of multiple data sources to estimate all quantities of interest. This review summarises the different types of stochastic epidemic models that use evidence synthesis and highlights current challenges.
\end{abstract}

\begin{keyword}
\kwd{evidence synthesis}
\kwd{state-space models}
\kwd{epidemic modelling}
\kwd{mechanistic modelling}
\end{keyword}

\end{frontmatter}

\section{Background}\label{sec:intro}
Epidemic models have become increasingly central to public health decision making, providing quantitative support to the efficient planning of health-care resources, the determination of optimal control strategies and the assessment of interventions to interrupt disease transmission. All of these require knowledge on hidden aspects of epidemics, such as current disease prevalence, severity, incidence and transmission, which can only been inferred through modelling. As a consequence of this crucial role of models, the methodologies underpinning epidemic modelling has come under increasing scrutiny. 
This has lead to more frequent adoption of rigorous approaches to linking models to data \cite{Heesterbeek2015ModelingHealth}, increasing model complexity and the need to use rich data arrays to guarantee reliable estimation. The result has been a recent proliferation of models incorporating data from multiple sources \citep[e.g.][]{AdeS06,Lio2013}.

In an attempt to summarise and critically review this literature, we will characterise models using a common construct. Most epidemic models can be expressed in terms of a general state-space framework:
\begin{align}
&&&&\vecX_t &\sim p_{\vphi}\left(\cdot\lvert\vec{x}_{t-1}\right)&&\text{(state equation)}\label{eqn:system}\\
&&&&\vecY_t &\sim p_{(\vphi, \veta)}\left(\cdot\lvert\vec{x}_t\right)&&\text{(observation equation)}\label{eqn:observation}
\end{align}
where $t = 1, \ldots, T$ and  the $p(\cdot\lvert\cdot)$ are appropriately chosen probability density functions \citep{Brockwell2002}. The state equation \eqref{eqn:system} governs the evolution of an epidemic system, represented by a state vector $\vecX_t$,  
characterised by a vector-valued parameter $\vphi$.
Equation \ref{eqn:observation} relates the underlying epidemic system to relevant data $\vecY_t$. These data are typically imperfect observations associated with $\vecX_t$, constrained by the limitations of surveillance schemes and subject to a nuisance parameter, $\veta$.
State vectors typically consist of all latent quantities that change over time (usually probabilistically), and $\vphi$ governs their temporal development. In some cases, the state vector is simply a function of $\vphi$. More commonly, epidemic models are compartmental, partitioning a population according to, for example, infection status. The numbers of individuals in each model compartment is included in the state vector, as is any quantity describing model dynamics that evolves over time, {\it e.g.} infection intensity \cite{Birrell2012} or the transmission potential \cite{Shubin2016}).

The focus of the statistical analysis could be to estimate unobserved system states $\vecX_{1:T}$ either sequentially (filtering) or retrospectively (smoothing), and/or to make inference about components of $\vtheta = (\vphi, \veta)$ that have some crucial interpretation. These parameter components might measure some headline statistic for the epidemic, such as the epidemic's reproductive number $R_0$, the average number of secondary infections caused by a single primary infection in a wholly susceptible population, or the effect of an intervention. 
This inference, ideally, would be based on direct observations on the epidemic system, {\it i.e.}
\begin{equation}\label{eqn:rw}
\vecY_t = \vecX_t + \veta^T \veps_{Y,t}, \text{ where } \veps_{Y,t} \sim N\left(\vec{0},\vec{I}\right).
\end{equation}
However \eqref{eqn:rw} implies observation of, for instance, new infections as they occur, which, especially in large populations, is rarely feasible.
More realistically, data are indirectly related to the quantities of interest and inference becomes possible only through the integration of data from multiple sources, allowing, for example, the evaluation of biases, separation of signal from noise or the interpolation of missing data.
Thus, given $\vtheta$, $\vecY_t$ is a collection of $N$ independent datapoints $(\vec{y}^1_t,\ldots,\vec{y}_t^N)$.

Evidence does not just come in the form of data. There are also modelling assumptions that underlie the parametric forms of $p_{\vphi}(\cdot)$ and $p_{(\vphi,\veta)}(\cdot)$, based on relevant literature, expert opinion and/or collateral data not included in the modelling process. In particular, pragmatic choices might need to be made over which parameter components can realistically be estimated by the available data, and which components it is prudent to assume to be known from literature. Synthesis of this kind of {\it a priori} evidence can be formalised by adopting a Bayesian framework centered on the posterior distribution
\begin{equation}\label{eqn:post}
p\left(\vtheta, \vec{x}_{1:T}\lvert\vec{y}_{1:T}\right)\propto p\left(\vec{y}_{1:T}\lvert\vec{x}_{1:T},\vtheta\right)p\left(\vec{x}_{1:T}\lvert\vtheta\right)
p\left(\vtheta\right),
\end{equation}
where $p(\vtheta)$, the prior distribution for $\vtheta$, encodes all that is known of $\vtheta$ from sources external to the present study. The posterior distribution represents a 
natural synthesis of this additional external information with $\vec{y}_{1:T}$.

In this paper, we shall provide an overview of evidence syntheses in stochastic epidemic modelling where multiple types of data are explicitly used in an integrated analysis. In Section \ref{sec:static} we will focus on non-mechanistic statistical models for epidemic data. Initially these models will be static, and the aim of the analysis is to estimate the current state of an epidemic. This set-up will then be extended by adding a time dimension, initially to estimate time-varying incidence. In Section \ref{sec:deterministic} we consider how multiple sources of data are used for inference in mechanistic models for disease transmission, where the dynamics governing transmission are assumed to be deterministic ({\it i.e.} $\var{\vecX_t\lvert\vec{x}_{t-1}} = 0, \forall t$), so that stochasticity is only provided by the observational component. Section \ref{sec:stochastic} reviews evidence syntheses in epidemic models with stochastic dynamics ({\it i.e.} $\var{\vecX_t\lvert\vec{x}_{t-1}} \neq 0$). The paper concludes with a discussion, identifying some ongoing and future challenges in the use of multiple datasets in stochastic epidemic modelling.

\section{Non-mechanistic Epidemic Modelling}\label{sec:static}

\subsection{Static Models}\label{sec:substatic}

Often estimation of the state of an epidemic at a particular point in time is of interest. In such examples, static or ``snapshot'' models are used, and the temporal evolution in Equations \ref{eqn:system} and \ref{eqn:observation} is not relevant:
\begin{align*}
\vecX &\sim p_{\vphi}\left(\cdot\right)\\
\vecY &\sim p_{\vtheta}\left(\cdot\lvert\vecX\right)
\end{align*}
In many cases, $\vecX$ will be a deterministic function of $\vphi$, {\it i.e.} $\vecX\equiv\vecX(\vphi)$, or can be integrated out of the analysis entirely if estimation of $\vphi$ is the focus. Therefore, for notational ease, we shall write $\vtheta = (\vphi,\veta,\vecX)$.

As in Section \ref{sec:intro}, come in the form of $N$ independent components $\vec{y} = (\vec{y}^1, \ldots, \vec{y}^N)$, where each $\vec{y}^n, n \in 1, \ldots, N$ may be multivariate. The aim of the evidence synthesis is to estimate a set of $K$ \emph{basic} parameters $\vtheta = (\theta_1, \ldots, \theta_K)$ from the complete array of information.  Each dataset $\vec{y}^n$ is assumed to inform a function $\psi_n = \psi_n(\boldsymbol{\theta})$ of the basic parameters, where $\psi_n$ is denoted a \emph{functional} parameter. If  $\psi_n(\boldsymbol{\theta}) \equiv \theta_k$, the data $\vec{y}^n$ are said to \emph{directly} inform $\theta_k$, whereas if the function is more complex and/or a function of multiple components of $\vtheta$, $\vec{y}^n$ \emph{indirectly} informs one or more parameters. Denote by $\vec{\psi}$ the collection of functional parameters $(\psi_1, \ldots \psi_N)$. Assuming conditional independence of each dataset, the likelihood is then
$$
L(\vtheta; \vec{y}) = \prod^N_{n=1} L_n(\psi_n(\vtheta); \vec{y}^n) 
$$
where each $L_n(\psi_n(\vtheta); \vec{y}^n)$ is the contribution of $\vec{y}^n$ to the basic parameters. This likelihood is either maximised, in a frequentist setting, or, in the Bayesian setting we consider here, a posterior distribution is obtained (equation \eqref{eqn:post}), summarising all information, both direct and indirect (as well as prior) on the basic parameters.

\begin{figure}
\includegraphics[width=0.3125\linewidth]{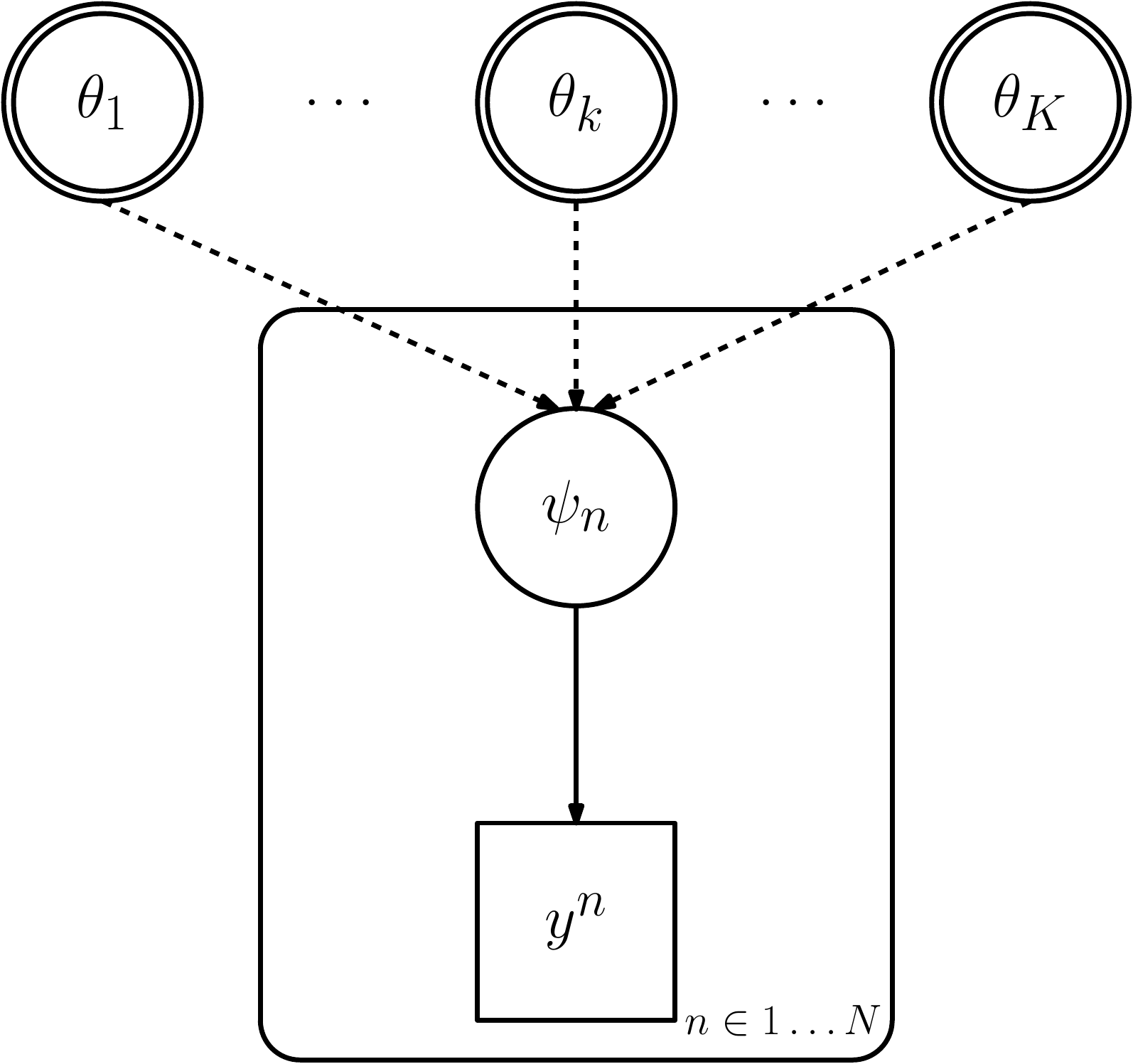}
\caption{Directed Acyclic Graph of a model with basic parameters, functional parameters and data.}\label{fig_evsynDAG}
\end{figure}

Such an evidence synthesis model can be represented as a directed acyclic graph (DAG) that encodes the conditional independence assumptions \citep{lauritzen1996graphical}. In the example of Figure \ref{fig_evsynDAG}, each basic parameter $\theta_k \in \vtheta$, denoted by double circles, is a \emph{founder} node of the DAG, \emph{i.e.} using family relationships to describe the relationships between nodes in the DAG, it has no parents nodes, only descendants. Functional parameters $\psi_n \in \vec{\psi}$ (single circles) are children of the basic parameters of which they are functions, with the dashed arrows denoting the (deterministic) functional relationship. By contrast, a solid arrow denotes a distributional (stochastic) relationship between nodes. Squares denote observed quantities $\vec{y}^n$. In a more complex hierarchical model with multiple levels, \emph{consequential} nodes internal to the DAG may be either deterministically or stochastically related to their ancestors or descendants. Repetition over variables is represented by `plates', rounded rectangles surrounding the repeated nodes, as for example the repetition of each $\vec{y}^n, n \in 1 \ldots N$ informing a different functional parameter $\psi_n$ in the figure.

\begin{figure}[!t]
\centering
\includegraphics[width=0.75\textwidth]{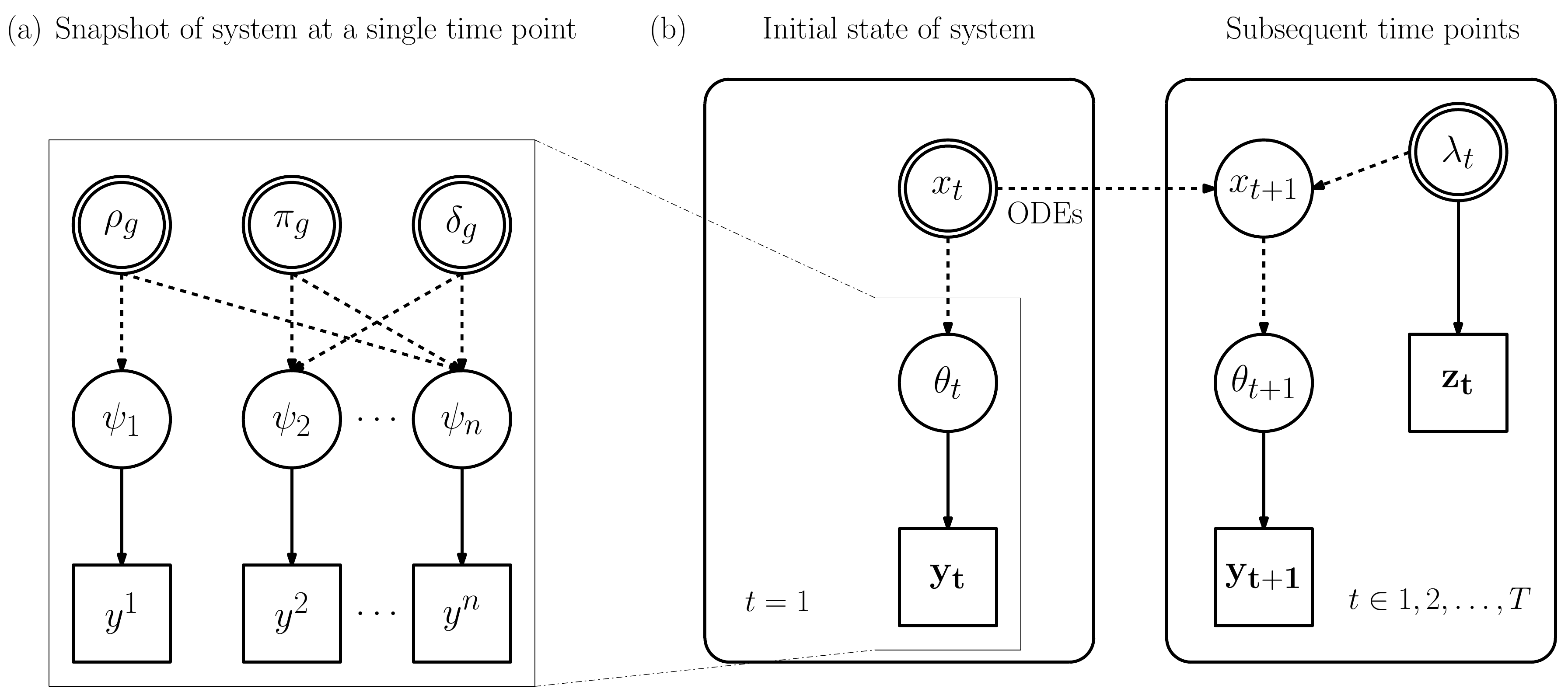}
\caption{(a) 
DAG of a HIV prevalence model with basic parameters $\vec{\theta} = (\rho_g, \pi_g, \delta_g)$. (b) Linking a series of snapshot HIV prevalence models at multiple time points $t$, to estimate HIV incidence in a ODE-driven compartmental model. Time $t$ data $\mathbf{y_t} = (y^1_t,y^2_t,\ldots,y^n_t)$ are augmented byby demographic and other transition rate data $\mathbf{z_t} = (y^{n+1}_t,\ldots,y^N_t)$. The parameters from (a), both basic and functional, are now encapsulated within $\vtheta_t$.}\label{fig:hivDAG}
\end{figure}

\vspace{-13.5pt} Evidence synthesis methods of this type in the context of medical and healthcare data were introduced in a synthesis of HIV prevalence data from different groups by \cite{Ades2002} and reviewed in \cite{AdeS06}. These have inspired a proliferation of comprehensive evidence syntheses particularly for static models of infectious disease, a prime example of which is 
the United Kingdom (UK) annual HIV prevalence (and, in particular, undiagnosed prevalence) estimates (\url{https://www.gov.uk/government/statistics/hiv-in-the-united-kingdom}). These are produced from multiple routine HIV surveillance datasets combined with contemporaneous cross-sectional survey data.
Figure \ref{fig:hivDAG}(a) presents a DAG of this general approach, 
summarised in \cite{DeAngelis2014a}. Here the  $\vec{\psi}$ are expressed as a function of basic parameters: $\rho_g$, the proportion of a population in a particular risk group $g$ for HIV;
$\pi_g$, the proportion of group $g$ infected; and $\delta_g$, the proportion of infections in group $g$ that are detected (diagnosed). 
This work has spread beyond the UK to the Netherlands \citep{Conti2011} and Poland \citep{Rosinska2016}.

Outside of HIV, similar analytical principles have been used to monitor the Hepatitis C virus (HCV) \citep{Sweeting2008a,DeAngelis2009}. In common with the HIV models, the prevalence of HCV in hard-to-reach populations, such as people who inject drugs (PWID), requires simultaneous estimation of the proportion of the population who are PWID 
 as well as HCV prevalence \citep{Harris2012,Prevost2015,McDonald2014b}. The estimation of influenza severity, measured by attack rates ({\it i.e.} cumulative incidence) and case-severity risks (probabilities of severe health events, such as hospitalisation, given infection) has also been approached through evidence syntheses \cite{PresanisEtAl2009,PresanisEtAl2011a,ShubinEtAl2013,Mcdonald2014}. Moreover, the healthcare burden from campylobacter infection has been similarly studied \cite{AlbertEtAl2011}.

Although the motivation behind evidence synthesis is to frame all the available information on the state of an epidemic within a single integrated analysis, for a number of reasons, including computational efficiency or uncertainty in model structure, it may be convenient to break the problem into smaller components. Whereas \cite{Prevost2015} fit a model for HCV prevalence in two stages, \cite{Price2016} present a synthesis of many evidence syntheses, in an attempt to characterise the link between chlamydia infection and long-term damage to reproductive health. In  \cite{PresanisEtAl2011a} three waves of \flu~infection in England are modelled in near isolation, with only weak correlation of the numbers infected in each wave. Although this `modular' approach is often reasonable and convenient, merging the resulting sub-models into a single analysis is non-trivial (see Section \ref{sec:discussion}). 

\subsection{Dynamic Models}\label{sec:dynamic}

In most cases, interest will lie in estimating the evolution of an epidemic, and in the rates of infection, in particular. One way to uncover this temporal variation is by linking snapshot analyses. At time $t$, data are distributed
\begin{equation*}
\vecY_t \sim p_{\vtheta}\left(\cdot\lvert\vecX_t\right),
\end{equation*}
and are linked over time via some smoothing of the state variables $\vecX_t$. This linkage is achieved for the HIV prevalence example (see Figure \ref{fig:hivDAG}(a)) by imposing a multi-state model \cite{PresanisEtAl2011}, partitioning the population into four disease states $\vecX_t$. This continuous-time model uses a system of ordinary differential equations (ODEs) to describe model dynamics.
Time-varying transition rates, including HIV incidence, become the piecewise-constant basic parameters $\vphi = \left\{\vec{\lambda}_g(t)\right\}$, the identifiability of which is feasible through the inclusion of additional demographic data $\vec{z}_t$.

A further example of snapshot-type evidence synthesis over time is in toxoplasmosis
\cite{Welton2005} where temporal smoothing is through random walk distributions placed on the log-incidence $\vecX_t$:
\begin{equation*}
\log\left(\vecX_t\right) \sim \log\left(\vecX_{t-1}\right) + \vphi^T\veps_{X,t}.
\end{equation*}
Using more routine time series type data, a similar random walk is used to smooth pertussis incidence in \cite{McdonaldEtAl2015}. Random walks are also used in a study of swine influenza prevalence in abattoirs \cite{StrelioffEtAl2013}, where, routine virological (swabbing to test for viral presence) and serological (testing of blood samples for an immunological response) data are used to distinguish recent from longer-standing infections. Here, $\vecX_t$ is composed of the proportion positive in the two data streams.

Back-calculation represents an approach to estimating incidence when the available data are time series counts of clinical endpoints. The classic back-calculation example is the estimation of HIV incidence from AIDS diagnoses \citep{BroG88}, based on the convolution equation:
\begin{equation}\label{eqn:convol}
\mu(t) = \int_0^t h(s) f(t-s) ds.
\end{equation}
Here $\mu(t)$ is the time-varying rate at which diagnoses occur, which is expressed in terms of an assumed known distribution function for the incubation period $f(\cdot)$ and the rate $h(\cdot)$ at which new infections occur. Versions of \eqref{eqn:convol} 
that use additional sources of information have been developed \cite{DeAGD98} and extended
to tackle the challenges posed by therapeutical advances and developments in HIV surveillance, augmenting AIDS diagnoses with HIV diagnoses \cite{backcalculation}. Multi-state versions have been proposed \citep{AalFDDG97}, with states defined by the levels of markers of HIV progression ({\it{e.g.}} CD4 cell counts), characterising the diagnosis process as the result of disease progression and the propensity to test. HIV diagnosis data have been further augmented in \cite{SweDA05} by observations on CD4 counts around diagnosis, to estimate the diagnosis process and the number of undiagnosed infections by disease state.
This work is extended in \cite{Birrell2012}, using random-walks to model the evolution of incidence and diagnosis rates over time \cite{WuTan2000}. Back-calculations of this type can be framed in terms of a state-space model by letting the state vector $\vec{X}_t$ include incidence $h_t$, diagnosis rates and state occupancies. The parameter $\vtheta$ is composed of initial conditions and random-walk variances. If the model is evaluated at discrete intervals, with no observation error assumed, then diagnoses are Poisson distributed and the CD4 data follow a multinomial distribution, falling into the following formulation: 
\begin{align*}
\vecX_t &= p_{\vphi}\left(\cdot\lvert\vecX_{t-1}\right)\\
\vecY_t &= g_{\vtheta}\left(\vecX_t\right) 
\end{align*}
The observation model becomes non-trivial in the presence of overdispersion.
Additional data from recent infection testing algorithms (RITA) to identify recent infections amongst newly diagnosed individuals are becoming increasingly available \citep{SomCLMA11} and preliminary attempts to integrate this type of data into a more traditional back-calculation have been made \cite{YanZW11}.
Inversely, in a study to determine the contribution of early HIV infections towards transmission, \cite{VolIRBMK13} fitted a back-calculation model in a first stage of analysis, before using HIV genetic sequence data, alongside other epidemiological information, to estimate the timing of HIV transmission.

\section{Evidence Synthesis in Mechanistic Transmission Models}\label{sec:deterministic}

The classic approach to tracking the spread of an epidemic is through compartmental models that partition the population into susceptible/infected/removed (SIR) categories \citep{Anderson1991} - or one of many similar variants. In the epidemic modelling literature these models 
are labelled as mechanistic transmission models. They differ from the multi-state models described earlier due to the explicit modelling of the transmission mechanisms, where rates of infection are a function of the prevalent number of infected and infectious individuals. 

The dynamics of such models unfold according to a system of ordinary or stochastic differential equations (or their discrete-time difference approximations). 
Initially, focus will be on models for which there is a deterministic state relation, but the epidemic is imperfectly observed. These can be expressed as:
\begin{equation}
\begin{split}\label{eqn:deterministic}
\vecX_t &= f_{\vphi}\left(\vecX_{t-1}\right)\\
\vecY_t &\sim p_{\vtheta}\left(\cdot\lvert\vecX_t\right).
\end{split}
\end{equation}
where $f_{\vphi}(\cdot)$ indicates a deterministic functional relationship, characterised by parameter $\vphi$ and $\vecX_t$ represents the number of people within each stage of the SIR-type model.
Typically, $\vphi$ will include rates of transition between model states, relative rates of contact between different population strata and the transmission potential. These movements between model states are unobserved, and, as in Section \ref{sec:dynamic}, the use of multiple data sources becomes necessary for identifiability. Several examples exist, where epidemic surveillance data are synthesised with: serological data; demographic, administrative or environmental data; and/or phylogenetic data.

\subsection{Surveillance and Serological Data}
Serological data identify the proportion of the population in the susceptible state. Typically these data are binomially distributed with the probability parameter at time $t$ given by
\begin{equation*}
\mathbb{P}(\text{seropositive at time }t) = 1 - S_t / N
\end{equation*}
where $S_t$ is the number of susceptibles among a population of size $N$. As an epidemic unfolds, the increase in the seropositivity of samples should mirror the levels of cumulative incidence. 

The work in \cite{BirrellEtAl2011} highlights the important role played by serology data in uncovering an epidemic's dynamics. Here, serological data are used to provide information on the scale of infection. Due to the presence of asymptomatic infection, this scale could not be derived while the epidemic is ongoing from data on general practice (GP) consultations for influenza-like illness and associated virological swabbing alone.
A similar approach is applied to data from Israel \cite{Yaari2016}, and \cite{DorigattiEtAl2013} extend this work to look at the changes in the immunity profile of a population and the fluctuating transmissibility of the virus between temporally distinct waves of infection.
Given the importance of serological data, \cite{teBeestEtAl2015} develops the approach further in application to Dutch \flu, taking into account the sensitivity and specificity of the serological testing process. The authors model 
actual titre values, giving
a probabilistic interpretation of immunity.

Outside of the 2009 pandemic, \cite{BaguelinEtAl2013} model individual `flu seasons on the basis of primary care data (consultations plus swab positivity and, where they exist, serological data), whilst also making some allowance for the uncertainty in the contact matrices that are used as a central fulcrum of the transmission model.

\subsection{Surveillance and Demographic, Administrative or Environmental Data}

An example of joint modelling of surveillance and demographic data is in \cite{PresanisEtAl2011}, where 
the model in Figure \ref{fig:hivDAG}(b) is extended to include a component of disease transmission
  utilising information $\vec{Z}_t = \vec{z}_t$ on ageing, migration and mortality. This is a rare example where such data contributes to the likelihood. More commonly demographic data are used as explanatory variables. 
  Here, assume that the observation model is defined for data $\vecY_t = \vec{y}_t$, but we also have explanatory data $\vec{Z}_t$. Then the system equation in \eqref{eqn:deterministic} is replaced by
  \begin{equation*}
\vecX_t = f_{\vec{\phi}}\left(\vecX_{t-1}, \vec{Z}_t\right).
  \end{equation*}
These explanatory data can come in many forms: \cite{BaguelinEtAl2013} and \cite{Shubin2016} use vaccination data to inform transition rates out of a susceptible state; \cite{Birrell2016} use commuting data to describe inter-region transmission; \cite{Yaari2016} relate transmission of \flu~in Israel to an index of `mean absolute humidity'. One particularly successful example of this type of data has been the use of air traffic data in the GLEAM system for the global tracking (and prediction) of a pandemic influenza outbreak \citep[e.g.][]{Tizzoni2012}. Here, the global spread of pandemic outbreaks is identified by short time series of virologically-confirmed \flu~cases, a spread that is assumed to occur along the lines of a network formed on the basis of established flight paths.

\subsection{Surveillance and Phylogenetic Data}\label{sec:phylo}

One synthesis of two very different types of information is evident where genetic analyses of pathogen genomes are used to provide or augment inference on the spread of the pathogen through a population. Gene genealogies can be reconstructed from gene sequence data to produce phylogenetic trees (or phylogenies). The branching points of the phylogenies can then be used to inform a simple coalescent model, the rate of coalescence of which can be linked to the number of infective individuals within the population.

Initially, phylogenies were used to estimate epidemic growth rates and epidemic emergence times using simple, non-mechanistic models under strict assumptions \citep{RasmussenEtAl2011}. However, in  \cite{pybus2001epidemic} and \cite{Volz2009PhylodynamicsEpidemics} there is a body of work emerging to link the phylogenies into mechanistic transmission models, with examples in HCV and HIV infection respectively. \cite{DearloveWilson2013} develop a methodology for phylodynamic models, showing that an SIR model can be identifiable on the basis of genomic data alone, without augmenting using surveillance data. \cite{RasmussenEtAl2011}, however, note that the use of phylogenetic information is of particular utility in the case where the surveillance counts are highly noisy or only weakly informative, and their work in links phylogenies to a continuous-time, continuous-space stochastic epidemic model informed by both epidemiological surveillance data and phylogenetic data.

\subsection{Stochastic epidemic models}\label{sec:stochastic}

In many cases, the deterministic dynamics are not sufficient. When numbers infected are sufficiently small that stochastic fluctuations in transmission can significantly impact on the future epidemic trajectory, deterministic epidemics can lead to over-optimistic forecasts, and can exclude the non-zero possibility of epidemic extinction when the epidemic reproductive number, $R$ is $>1$. Similarly, when the epidemic unfolds in the presence of environmental or other external factors not typically captured by the transmission model, stochasticity in the temporal evolution of parameter values can have a significant impact on the pattern and timing of infections. In either case, these models require the full state-space specification of equations \eqref{eqn:system} and \eqref{eqn:observation}.

In the case of demographic stochasticity, chain-binomial \citep{LekF06}, chain multinomial \citep{WuTan2000} or chain negative binomial \citep{Finkenstadt2002} models may be used. It is, however, the second context of unexplained variablility that is more prevalent in the literature. 
In particular, the transmission potential $\beta(t)$ is commonly modelled as a stochastic process. \cite{DurKB13,Xu2016a} cast $\beta(t)$ as Wiener and Gaussian processes respectively, whereas \cite{Shubin2016} impose a random effects model on $\beta(t)$.

Though the motivation for the use of multiple sources of data in stochastic epidemic modelling is frequently no different to the deterministic case, there are few examples of their use in the literature. Using traditional forms of influenza surveillance, \cite{Shubin2016} constitutes a rare example, where laboratory-confirmed data on `mild' cases are combined with data on (nested) admissions to hospitals and to ICUs. Both types of stochasticity are in effect also, as the model is in discrete time and assumes chain binomial transmission, whilst the force of infection arises as the result of a random effects model. 
Stochastic epidemic models have been used in analyses of the type discussed in Section \ref{sec:phylo}, initially through the modelling study of
\cite{RasmussenEtAl2011}, who propose an SIR model allowing for both seasonality and environmental noise in transmission. This work is extended by \cite{RatDMFK12} through a case study of competing viral subtypes of A/H3N2 influenza, whose transmission model incorporates demographic (chain-multinomial) stochasticity, fitting the overall phylodynamic model using approximate Bayesian methods.

\section{Discussion}\label{sec:discussion}

The recent increase in the number of evidence syntheses, mostly Bayesian, to estimate latent characteristics of epidemics is testimony of the crucial role of data from multiple sources. This role has been comprehensively explored in other reviews \citep{AdeS06, DeAngelis2014a}, but include the ability to: identify and estimate key quantities that are not directly observed; introduce and formally quantify expert judgement in the form of prior distributions; readily account for and estimate known biases in observational data through the introduction of bias parameters with carefully chosen priors; and the minimisation of selection bias through the use of all available relevant data, both direct and indirect. 

However, although the adoption of evidence synthesis methods signal an ability to fit models of increasing realism and complexity, some general challenges remain \cite{DeAngelis2014} and are the subject of active research.

\subsection{Model building}

As briefly discussed in Section \ref{sec:substatic}, it may be convenient to divide a complex model into smaller, more manageable sub-models. Reasons include computational tractability as well as the need to explore alternative model formulations during a model development phase. 
However, summarising the results of each sub-model into a second-stage ``full'' model in a manner that retains the feedback from different data sources to common parameters is not straightforward. Recent work that allows for principled inference from a fully joint model given posterior samples from each sub-model has been proposed \citep{Goudie2016r}. The application of this ``Markov melding'' approach to evidence syntheses in the stochastic epidemic field has the potential to address some of the computational challenges that arise from complex stochastic models.

\subsection{Model criticism}

In any model-based analysis, critical assessment is important, but there are some particular aspects that arise in multiple source analyses in the context of epidemic modelling. Since so many epidemic characteristics are not observable directly, the question of which parameters are identifiable (partially or fully) from the available data sources is an active area of research \citep{Gustafson2010a}. How to determine algebraically, ahead of any inference, which parameters are potentially identifiable in a complex dynamic system has been explored recently in the systems biology field \citep[e.g.][]{Gross2016}: such methods have the potential to be adapted to transmission models. An alternative is to consider adapting value-of-information methods \citep[e.g.][]{ParmigianiInoue2009} to the case of evaluating gains in precision in parameter estimates resulting from collecting or incorporating further evidence.

In contrast to weak identifiability issues, multiple data sources may inform the same parameter, leading to the potential for conflicting evidence \citep{DeAngelis2014}. Such potential represents a further motivation for modular model-building, uncovering the influence of each additional data source. As already discussed however, such modular approaches are computationally intensive, so the use of cross-validatory conflict diagnostics, both at the model-building and model-assessment stages, requires adaptation to enable timely inference.

Once conflicts have been detected and measured, they require resolution. Typically, conflicts arise as a result of unaccounted bias and/or naive interpretation of what the data represent. Bias modelling approaches to resolve conflict are typically related to ideas of how best to weight different sources of evidence in a synthesis \citep{DeAngelis2014}, which is still an open question. While in a frequentist framework there are well-established methods to account for selection biases in the types of observational data usually included in epidemic evidence syntheses, Bayesian equivalents are still in their infancy \citep{Si2013}. 

\subsection{Efficient inference}
The other key challenge area is that of computationally efficient statistical inference. This is of particular importance in the context of epidemic modelling, not only as a way of addressing the complexity inherent in a realistic model of a stochastic process informed by multiple data sources, but also because real-time estimation is crucial for addressing public health policy needs in the midst of an emerging epidemic \citep{DeAngelis2014}. Much progress has been made in developing and applying efficient algorithms for epidemic evidence syntheses, such as: sequential Bayesian methods \citep{Sheinson2014ComparisonEpidemic,Birrell2016EfficientFeasible}, including likelihood-free particle MCMC \citep{RasmussenEtAl2011}; approximate Bayesian computation \citep{RatDMFK12}. Alternatively, to achieve efficient inference, one might approximate the complex epidemic model with a readily implementable proxy. Shaman and colleagues have produced a number of papers \citep[e.g.][see also \cite{Simons2012} for an application to global measles data]{ShaKYTL13} that use an extended Kalman filter (EKF) to provide a stochastic time series approximation to the dynamics of SIR models, whereas Bayesian emulation \citep{Farah2014} seeks to characterise an epidemic model (or simulator) with an emulator, built from a dynamic Gaussian process prior. The next challenge is to broaden the scope of such algorithms to handle multiple datasets, possibly diverse in nature.

\subsection{Conclusions}
A recent review of infectious disease modelling \cite{Lessler2016} 
suggests that the full potential of mechanistic models that ``simultaneously link data from diverse, heterogeneous data sources'' has yet to be reached. 
This is certainly true for fully stochastic transmission models, though rare examples do exist \citep{RatDMFK12,Shubin2016} of such models embedded within an evidence synthesis. Such rarity and the challenges discussed above motivate the need for further development in this area. 
Phylodynamic modelling, in particular, offers a very natural application, relying as it does upon very different types of data.
Here, approaches to propagating the uncertainty in the ascertainment of phylogenies into models for transmission, and to handling multiple exposures and the presence of within-host phenotypic variation all present significant methodological challenges.

However, the many examples reviewed in Section \ref{sec:deterministic}, particularly for deterministic models, suggest that evidence synthesis for mechanistic models is both a well-established and rapidly expanding field.

\bibliographystyle{imsart-number}
\bibliography{Mendeley_SSSES-paper-PB}

\end{document}